\documentclass{article}

\usepackage{microtype}
\usepackage{graphicx}
\usepackage{subfigure}
\usepackage{booktabs} 
\usepackage{hyperref}
\usepackage{amsmath}

\usepackage[accepted]{icml2019}

\icmltitlerunning{Quarantine Deceiving Yelp's Users by Detecting Unreliable Rating Reviews}

\begin{document}

\twocolumn[
\icmltitle{Quarantine Deceiving Yelp's Users by Detecting Unreliable Rating Reviews}



\icmlsetsymbol{equal}{*}

\begin{icmlauthorlist}
\icmlauthor{Viet Trinh}{equal,institution}
\icmlauthor{Vikrant More}{equal,institution}
\icmlauthor{Samira Zare}{equal,institution}
\icmlauthor{Sheideh Homayon}{equal,institution}
\end{icmlauthorlist}

\icmlaffiliation{institution}{Department of Computer Science and Engineering, Baskin School of Engineering, University of California, Santa Cruz, CA, USA}

\icmlcorrespondingauthor{Viet Trinh}{vqtrinh@ucsc.edu}
\icmlcorrespondingauthor{Vikrant More}{vmore.edu}
\icmlcorrespondingauthor{Samira Zare}{szare@ucsc.edu}
\icmlcorrespondingauthor{Sheideh Homayon}{shomayon@ucsc.edu}


\vskip 0.3in
]



\printAffiliationsAndNotice{\icmlEqualContribution} 

\begin{abstract}
Online reviews have become a valuable and significant resource, for not only consumers but companies, in decision making. In the absence of a trusted system, highly popular and trustworthy internet users will be assumed as members of the trusted circle. In this paper, we describe our focus on quarantining deceiving Yelp’s users that employs both review spike detection (RSD) algorithm and spam detection technique in bridging review networks (BRN), on extracted key features. We found that more than 80\% of Yelp's accounts are unreliable, and more than 80\% of highly rated businesses are subject to spamming.
\end{abstract}

\section{The Problem Statement}
\begin{figure}[h]
	\begin{center}
	\centerline{\includegraphics[width=\columnwidth]{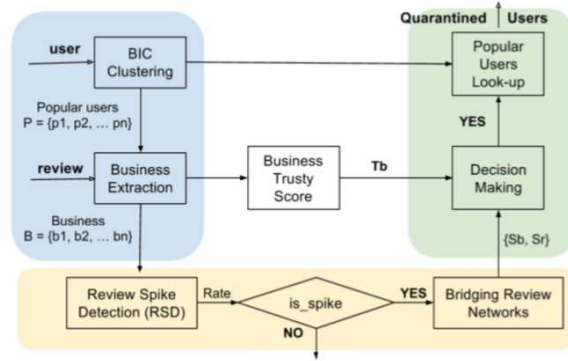}}
	\caption{The flowchart of quarantine deceiving Yelp users from Yelp’s user and review data set. The blue region denotes features extraction, the yellow region is responsible for detecting anomalous ratings, and the green region is for analysis and decision making in quarantining users. {\bf Tb} and  {\bf Sb}  denote a trusty score and a spam score for a business, and  {\bf Sr} is the calculated spam score for each review.}
	\label{fig:flowchart}
	\end{center}
\end{figure}

Online reviews, nowadays, have become a valuable and significant resource, for not only consumers but companies, in decision making. In the absence of a trusted system, highly {\em popular and trustworthy} internet users will be assumed as members of the trusted circle. The problem statement we are arguing is that: given a set of user rating reviews, determine whether they are trustworthy or unreliable; from these deceptive reviews of particular target business, identify and quarantine any Yelp account producing such over-threshold amount of those. Intrigued by the works of \cite{pranata2016most}, the figure \ref{fig:flowchart} visualizes an entire process, and the road map for tackling the learning problem is as follows:
\begin{itemize}
	\item Cluster the most popular Yelp’s users, and find businesses that these have rated
	\item Employ RSD algorithm \cite{rahman2015catch} and BRN technique \cite{rayana2015collective} to detect any unusual rating, analyze its corresponding business, and determine its validity
	\item Compare the popular users' disputable ratings, for those target businesses, against the business trusted values; and then, quarantine deceptive users based on comparison results
\end{itemize}

The rest of this report is constructed as follows: the next section is to discuss features selection for determining popular users, and potential spam scores of both deceptive reviews and target businesses; the third section introduces our hybrid mechanism, employing both RSD and BRN, to detect the review spike aiming to boost up a business’s goodness; the last sections mainly discuss the evaluation metrics, experimented results, and our final thoughts about this work.
\vspace*{-\baselineskip}
\vspace*{-\baselineskip}

\section{Features Extraction Engineering}
Since the number of users are large, the most efficient way to find those popular ones is to use clustering techniques that take a feature vector as its input. Similarly, much can be inferred from textual and behavioral data of reviews. This section outlines meaningful features extraction for serving the purpose of our work.

\subsection{Popular Users}
To cluster a group of most popular users out off nearly 650K records, we decide to use the {\em k-mean algorithm} with an input as a set of selected user features. The Yelp user records dataset provides a number of features to work with, as inputs to the k-mean algorithm; however, not all are useful. The following features are believed to be important, as they will have a significant impact on determining the most popular users:  {\bf yelping since} - the year in which an user started Yelp, {\bf average star} - the average ratings of all businesses given by an user, {\bf elite count} - the total number of years in which an user is part of the Yelp’s elite squad, {\bf fans count} - the total number of fans that an user has, {\bf friends count} - the total number of friends of a user, {\bf reviews count} - the total number of reviews that a user has made, {\bf total votes} -the total number of votes from others, and {\bf total compliments} -the total numbers of compliments from other users.

\subsection{Potential Fraudulent Reviews}
Textual features are derived from the reviews written by the users; whereas the behavioral features are computed from meta-data such as a review time stamp, ranks, and review patterns. We extract the following features for determining possible deceptive reviews: {\bf RD} - rating deviation, the absolute deviation of a rating from the businesses average values, as spammers tend to write unreliable reviews deviating from this value; {\bf EXT} - extremity of a review of 1 if a rating is 4,5 and 0 if it is 1,2,3; {\bf ETF} - the fact that a spammer wrote early reviews to make an impact on the overall rating of the business; {\bf ISR} - a value of 1 if the re- view is the user’s only review, otherwise 0; {\bf PCW} - the ratio of all capital words, as spammers tend to use a lot of capital words for drawing attention; {\bf PP1} - the ratio of the first person pronouns in reviews, as in deceptive ones, the second person pronouns are mostly used instead \cite{rayana2015collective}; {\bf EXC} - the number of exclamations used in the review to attract attention.

\subsection{Target Spamming Businesses}
Similarly, to calculate the potential spam score for each target business, the following features are extracted: {\bf MNR} - the maximum number of reviews written in a day for a business; {\bf PR, NR} - the ratio of positive reviews and negative reviews, as argued in \cite{rayana2015collective} that spammers rate more than 80\% of reviews as 4 or 5; {\bf avgRD} - the average rating deviation of business reviews that determines whether a business is being a target of spammed; {\bf ERD} - the entropy rating distribution of businesses reviews to determine the uncertainty of the distribution of review ratings; {\bf ETG} - the entropy of temporal gaps capturing any bursts in review activities; {\bf RL} - the average length of a review.

\section{Detecting Anomaly in Ratings}
\subsection{Clustering Popular Users}
The algorithm searches over a range of k values and select the best k clusters based on the approximation of the posteriors of the clusters, or Bayesian Information Criterion (BIC) score \cite{pranata2016most}. BIC is measured through the likelihood of how well the clusters model the data, which is produced by a spherical Gaussian distribution. Hence, the higher value of BIC is, the more probable of the clustering being a good fit is:
\begin{equation}
	BIC(D\: |\: k) = l(D\: |\: k) - \frac{p_j}{2}log(R)
\end{equation}

where $D$ is the dataset, $k$ is the number of clusters, $R$ is the number of feature vectors, $p_j$ is the number of parameters to estimate clustering, and $l(D\: |\: k)$ is the likelihood that is measured as:
\begin{multline}
	l(D\: |\: k) =  \sum_{i=1}^{k} - \frac{R_i}{2}log(2\pi) - \frac{R_id}{2}log(\delta^2) \\
	               - \frac{R_i - 1}{2} + R_i log(\frac{R_i}{R})
\end{multline}

where $R_i$ is the number of feature vectors in $i^{th}$ cluster, $d$ is the cluster center, and $\delta^2$ is the average variance of Eucledian distance from data point to its cluster center. Given a group of popular Yelp users $\mathcal{P}$, the Algorithm \ref{alg:biz-extraction} shows the steps taken to find the list of distinct businesses $\mathcal{B}$ that $\mathcal{P}$ has rated.
\begin{algorithm}[h]
	\caption{Business Extraction}
   	\label{alg:biz-extraction}
	\begin{algorithmic}
   		\STATE {\bfseries Input:} a group of popular users $\mathcal{P}$, review records $\mathcal{R}$
		\STATE {\bfseries Output:} list of distinct businesses $\mathcal{B}$ that $\mathcal{P}$ has rated
   			\FOR{$i=1$ {\bfseries to} $|\mathcal{P}|$}
				\STATE $\mathcal{B}_i \leftarrow $ FindBusiness($i$, $\mathcal{R}$)
				\FOR{$b=1$ {\bfseries to} $|\mathcal{B}_i|$} 	
					\IF{$\mathcal{R}_b \geq 10$ {\bfseries and} b $\not\in \mathcal{B}$} 
						\STATE $\mathcal{B} \leftarrow b$
					\ENDIF
				\ENDFOR
   			\ENDFOR
	\end{algorithmic}
\end{algorithm}

\subsection{Review Spike Detection (RSD)}
Deceptive users increase the rating of a target business by posting fraudulent reviews. The RSD module detects this by identifying {\em outlier} in a rating timeline that receives a higher number of positive (or negative) reviews than normal. Using the dispersion measurements of Box and Whiskers plots \cite{rahman2015catch,tamhane2000statistics,prithivirajan2015analysis} in detecting outliers, RSD computes the Upper Outlier Fence (UOF) value. The spike is detected when $|P_d| > $ UOF, where $P_d$ is a set of positive reviews posted during a day. For example, consider a business has an UOF value of 10, we say that there exists a positive spike if that business receives more than 10 positive reviews in any particular day. Similarly for detecting negative spikes.

\begin{figure}[h]
	\begin{center}
	\centerline{\includegraphics[width=\columnwidth]{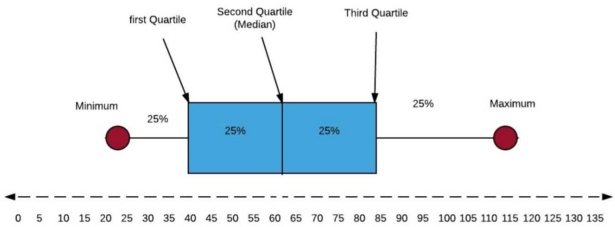}}
	\caption{Box and Whiskers plot for a review distribution over 125 days.}
	\label{fig:boxandwhiskers}
	\end{center}
\end{figure}

Fraudulent reviews can be posted either constantly in a short period of time or gradually over a long time interval. Thus, a sense of review distribution over time is necessary for determining a spike. According to  \cite{tamhane2000statistics}, the Box and Whiskers plot (figure~\ref{fig:boxandwhiskers}) yields five important statical summaries, in term of review-posted days: the most outdated, first quartile, second quartile or median, third quartile, the most current. For each business timeline, the interquartile range (IQR) is the difference between the first and third quartiles. In terms of ratings, Upper Outlier Fence (UOF) and Lower Outlier Fence (LOF), which have an abnormal distance from the others, must satisfy:
\begin{align}
	UOF \geq& Q3 + 1.5 (IQR)\\
	LOF \leq& Q1 - 1.5 (IQR)
\end{align}

where $IQR = Q3 - Q1$. A spiky business is detected when it has a set of reviews either exceeding UOF or falling under LOF.

\vspace*{-\baselineskip}
\vspace*{-\baselineskip}

\subsection{Bridging Networks for Calculating Spam Scores}
It is trivial that an extreme high (H) or low (L) rating is more suspicious for each target business. To quantify the extremity of a feature rating $r_i$, the proposed method from \cite{rayana2015collective} uses the empirical cumulative distribution function (CDF) to estimate the probability that a target business $\mathcal{B}_j$ contains a rating as low or as high as $r_i$. Specifically, for each rating $r_i$, $1 \leq r_i \leq |\mathcal{R}_b|$, CDF computes the $f$ value for the {\bf L} type by finding all ratings smaller than a target business' rating. Similarly, finding all values that are as large as a target one for computing the $f$ value of the {\bf H} type.
\begin{equation}
	f(r_{bi}) = \begin{cases} 
				1 - P(X \leq r_{bi}) &\mbox{if } H \\
				P(X \leq r_{bi}) &\mbox{if } L \\
			\end{cases}
\end{equation}

Both feature types eventually receive a low value if they are both suspicious. To get a stronger representation, we now combine all these values into a single spam score on a scale of 0 to 1 using:
\begin{equation}
	S = 1 - \sqrt{\frac{\sum_{i=1}^{|\mathcal{R}_b|} f(r_{bi})^2}{|\mathcal{R}_b|}} > S_{threshold}
\end{equation}

A higher value of $S$ indicates an abnormal review, or a business potentially being a target of spam. Generally, detecting a spike from RSD and evaluating its deceptive score will help in narrowing down a set of popular users $\mathcal{P}$ who manipulates a business $\mathcal{B}_b$'s rating.

\section{Evaluation}
In this report, we experiment on different size $k$ of clustering popular users. \cite{pranata2016most} performs clustering from $k = 2$ to $k = 60$ and find that $k = 4$ yielded the optimal group of popular users. However, our approach finds that the number of clusters does not affect an outcome of the popular users cluster. More specifically, we are able to determine 46 popular users, independently on how many clusters are being calculated, from a set of nearly 650K Yelp’s account. Details on this popular group will be provided in the next section.

Comparing against \cite{rahman2015catch}’s execution of RSD on an entire Yelp’s review dataset for finding spiky businesses, our input domain is much more smaller, hence more efficient. Since our approach starts with clustering popular users and their rated business, the RSD algorithm only needs to run over this set of popular businesses. Interestingly, despite of a difference in the review space, our finding agrees with \cite{rahman2015catch} that one will need to generate at least $\frac{n}{7}$ fraudulent reviews to increase the ratings of a business by a half point, where $n$ is the number of reliable reviews. The mathematical proof for this can be found in \cite{rahman2015catch}.

According to \cite{rayana2015collective}, a spam score $S$ for each business or a review is calculated to determine the probability of being a target for spamming. Similarly, we use this metric to determine whether a detected spike is deceptive. We argue that although a popular user might give an extreme {\em outlier} review, but it could be an honest one, due to personal interest or emotional feelings. Yelp’s users are only quarantined when they exceeds a limit of {\bf {\em deceptive}} reviews on a spiky business, and their reviews are far off a tolerated range of the business’ trusted score. \cite{pranata2016most} define a business’s trusted score as an average of all ratings; whereas we consider it to be an average of all {\bf {\em non-deceptive}} ratings: $T_b = \frac{1}{|\mathcal{R}_b|}\sum r$, where $b \not\in \mathcal{B}$.

\section{Results}
From our experiment, we find that there exists a group of 46 popular users affecting an entire review dataset significantly, regardless to how may clusters are being classified. Table \ref{tab:k4} and \ref{tab:k3} list all feature values of such clustering. It is noticeable that Cluster 2 seems to be the most popular one as all of its feature values are larger than the others. Our finding is some sort similar with \cite{pranata2016most}, as their work also confirms that Cluster 2 groups the most popular users; except it has 43 users. One feasible explanation is that we randomly select a set of k centroids for k-mean clustering, in which each feature is also randomly generated from a given range; whereas \cite{pranata2016most} specify particular centroids. Additionally, given this list of 46 popular users, 3202 businesses are founded by the algorithm \ref{alg:biz-extraction}.
\vspace*{-\baselineskip}
\begin{table}[h]
    \caption{Clustering results for Yelp's user records when $k = 4$}
    \label{tab:k4}
	\begin{center}
	\begin{small}
	\resizebox{0.85\columnwidth}{!}{
	\begin{tabular}{lccccr}
		\toprule
		Features & Cluster\_0 & Cluster\_1 & Cluster\_2 & Cluster\_3 \\
		\midrule
		yelping\_since & 2008.25 & 2008.58 & 2008.89 & 2012.38 \\
		average\_star & 3.88 & 3.84 & 3.85 & 3.75 \\
		elite\_count & 38.73 & 38.34 & 37.74 & 2.86\\
		fans & 390.22 & 262.35 & 548.76 & 1.18\\
		friends\_count & 20946.6 & 19874.26 & 30833.41 & 159.88 \\
		review\_count & 1437.97 & 1119.04 & 1994.20 & 25.35 \\
		total\_votes & 43021.45 & 28782.12 & 75479.22 & 83.55 \\
		total\_compliments & 21382.02 & 12293.42 & 19308.48 & 9.03 \\
		total\_users & 60 & 95 & 46 & 686355 \\
		\bottomrule
	\end{tabular}}
	\end{small}
	\end{center}
\end{table}
\vspace*{-\baselineskip}
\vspace*{-\baselineskip}

\begin{table}[h]
    \caption{Clustering results for Yelp's user records when $k = 3$}
    \label{tab:k4}
	\begin{center}
	\begin{small}
	\resizebox{0.9\columnwidth}{!}{
	\begin{tabular}{lcccr}
		\toprule
		Features & Cluster\_0 & Cluster\_1 & Cluster\_2 \\
		\midrule
		yelping\_since & 2008.25 & 2008.39 & 2008.89 \\
		average\_star & 3.88 & 3.74 & 3.85 \\
		elite\_count & 38.73 & 2.87 & 37.74 \\
		fans & 390.22 & 1.22 & 548.76 \\
		friends\_count & 20946.6 & 162.61 & 30833.41  \\
		review\_count & 1437.97 & 25.50 & 1994.20 \\
		total\_votes & 43021.45 & 87.25 & 75479.22  \\
		total\_compliments & 21382.02 & 10.73 & 19308.48 \\
		total\_users & 60 & 686450 & 46 \\
		\bottomrule
	\end{tabular}}
	\end{small}
	\end{center}
\end{table}

\begin{figure}[h]
	\begin{center}
	\centerline{\includegraphics[width=0.85\columnwidth]{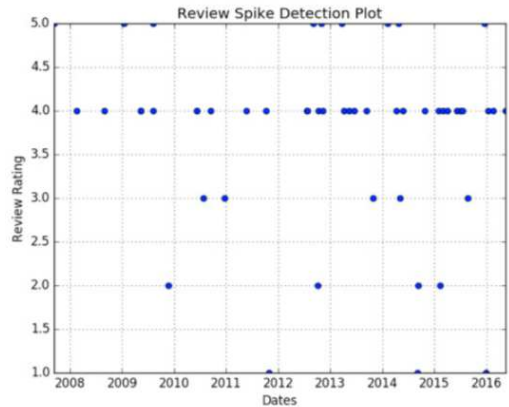}}
	\caption{A review spike detection plot for a business from 2008 - 2016. During the years from 2013 to 2015, this business receives a high amount of positive ratings whose values are larger than 3.0, mostly 4.0.}
	\label{fig:spike}
	\end{center}
\end{figure}

Moreover, we find 2715 spiky businesses which is 84.79\% of businesses rated by popular users, and this is less than the results of \cite{rahman2015catch}. It is expected as we only consider those rated by popular users; whereas \cite{rahman2015catch} take every single business into their account. Additionally, due to the fact that most of businesses has deceptive ratings spreading over a long period of time, the RSD algorithm runs on an entire timeframe of 2004 - 2016. The figure \ref{fig:spike} shows a sample RSD plot for a single business from 2008 to 1016.

\cite{pranata2016most} claim that most popular users are untrustworthy but do not provide any statistical results. In contrast, we try to understand how many popular users are actually untrustworthy due to the fact that, sometimes, their judgement is affected by emotional feelings. Since the rating range is from 1 to 5, we set the tolerate offset to be 0.5. This means that, given a trusted business score, ones will be classified as deceptive if their ratings are greater than $T_b + 0.5$ or less than $T_b - 0.5$. Additionally, a quarantined user must make more than a threshold amount of such ratings. The table \ref{tab:threshold} shows the correlation between a threshold amount of deceptive ratings and a number of quarantined users. Obviously, the smaller amount of threshold, the higher the percentage of quarantined users. In fact, it is reasonable to conclude that given the Yelp’s user dataset, there are 46 popular users and more than 80\% of those are untrustworthy.
\vspace*{-\baselineskip}
\begin{table}[h]
    \caption{The number of quarantined users based on the threshold amount of deceptive ratings}
    \label{tab:threshold}
	\begin{center}
	\begin{small}
	\resizebox{0.75\columnwidth}{!}{
	\begin{tabular}{lccr}
		\toprule
		Threshold & Number of Quarantine Users & Percentage \\
		\midrule
		3 & 43 & 93.48\% \\
		4 & 42 & 91.30\% \\
		5 & 42 & 91.30\% \\
		6 & 42 & 91.30\% \\
		7 & 41 & 89.13\% \\
		8& 39 & 84.78\% \\
		9 & 38 & 82.61\% \\
		10 & 38 & 82.61\% \\
		\bottomrule
	\end{tabular}}
	\end{small}
	\end{center}
\end{table}
\vspace*{-\baselineskip}

\section{Conclusion}
This paper aims to quarantine Yelp’s users making a large amount of deceptive reviews. Our approach starts with clustering popular users and their rated businesses. Employing RSD and BRN, we determine whether a detected spike for a business is trustworthy. A user is classified as deceptive when making an exceeding amount of reviews on a spiky business, whose values are out off a tolerated range. Conclusively, there are 46 popular users out of approximately 650,000 Yelp’s accounts, and more than 80\% of those are deceptive. Additionally, out of almost 3300 businesses rated by popular users, 2715 or 84.79\% of those are subject to spamming; and a popular deceptive user needs to generate at least a-seventh of a number of positive reviews for increasing a business general ratings by a half-star.

\nocite{langley00}

\bibliography{main}
\bibliographystyle{icml2019}

%
%
%

\end{document}